\documentclass[prd,preprint]{revtex4}
\usepackage[dvips]{color}
\usepackage[utf8]{inputenc}

\usepackage{latexsym}
\usepackage{epsfig}
\usepackage{graphicx,subfigure}
\usepackage[normalem]{ulem}
\usepackage{color}
\usepackage{xcolor}
\usepackage{multirow}
\usepackage{hyperref}
\usepackage{numprint}
\usepackage{eucal}
\usepackage{amsmath}
\usepackage{amssymb}

\usepackage{xcolor}

\newcommand{\Lg}{\mathcal{L}}

\newcommand{\h}{\mathcal{H}}

\newcommand{\ssout}[1]{}

\def\ga{\mathrel{\raise.3ex\hbox{$>$\kern-.75em\lower1ex\hbox{$\sim$}}}}
\def\la{\mathrel{\raise.3ex\hbox{$<$\kern-.75em\lower1ex\hbox{$\sim$}}}}

\def\be{\begin{equation}}
\def\ee{\end{equation}}
\def\bea{\begin{eqnarray}}
\def\eea{\end{eqnarray}}

\def\T{\rm T}

\def\L{\rm L}

\begin{document}

\author{Antonio Enea Romano}

%\affiliation{Theoretical Physics Department, CERN, CH-1211 Geneva 23, Switzerland}
\affiliation{Instituto de Fisica,Universidad de Antioquia,A.A.1226, Medellin, Colombia}
\affiliation{ICRANet, Piazza della Repubblica 10, I--65122 Pescara, Italy}

\title{The effects of dark energy on the matter-gravity coupling} 
\begin{abstract}
Dark energy perturbations are expected to modify the evolution of cosmological perturbations, producing different observational effects. We show that these effects can be encoded in a momentum, space and polarization dependent effective matter-gravity coupling. 
For scalar perturbations the effective gravitational coupling can be locally negative, in regions with local dark energy under-densities.
For adiabatic perturbations the effective gravitational coupling for scalar perturbations can be negative for phantom dark energy, providing a possible explanation for the observed structure suppression at low redshift, consistent with other independent evidences of evolving dark energy. For gravitational waves the effective coupling can make the radiation emitted by compact binary convalescences polarization and frequency dependent. 
\end{abstract}

% Dark energy perturbations are expected to modify the evolution of cosmological perturbations, producing different observational effects. % on scalar perturbations and gravitational waves. 
% We show that these effects can be encoded in a momentum, space and polarization dependent effective matter-gravity coupling. %, in agreement with the results obtained in some specific theories.
% %Using appropriately defined effective energy tensors, we show that the method can be extended to include also the effects of higher order perturbations.
% %The generality of this effective description of the effects of dark energy on cosmological perturbations gives a  theoretical justification for the phenomenological approaches used in the analysis of cosmological observations ,  allowing to understand the mapping between the results of the phenomenological analysis of observational data and theoretical models. 

% For scalar perturbations the effective gravitational coupling can be locally negative, in regions with local dark energy under-densities.
% For adiabatic perturbations the effective gravitational coupling for scalar perturbations can be negative for phantom dark energy, providing an explanation for the observed structure suppression at low redshift, consistent with other independent evidences of evolving dark energy.

\maketitle

\section{Introduction}
Dark energy accounts for most of the Universe energy density today, but its fundamental physical nature is still mysterious. For this reason modified gravity theories have been proposed as dark energy models, such as for example Horndeski gravity \cite{Horndeski:1974wa}, or other extensions \cite{Gleyzes:2014qga}.
In these theories the field equations are modified with respect to general relativity (GR), and the new equations can be conveniently re-cast in the form of Einstein equations with a running Planck mass and  an effective stress-energy tensor associated to the new fields and geometrical terms, which is interpreted as an effective  dark energy stress-energy tensor.
This effective dark energy tensor has perturbations, which modify the GR cosmological perturbations equations, introducing new scale and time dependent effects. We show that these effects can be encoded in different effective time, momentum and polarization dependent matter-gravity couplings.
This implies for example that the gravitational waves (GW) energy radiated by a binary system can have an additional  frequency and polarization dependency, due to the interaction with dark energy.

For scalar perturbations  the effective gravitational coupling  can be locally negative in regions with local dark energy underdensities. For scalar
adiabatic perturbations the scalar effective gravitational coupling can be negative for background phantom dark energy \cite{DESI:2024mwx}, providing a possible explanation for the observed structure suppression at low redshift \cite{Chen:2024vuf}, and a consistency test between the background and perturbations effects of dynamical dark energy.

\section{Modified gravity theories}

In order to formulate a general theoretical framework let's consider a  theory defined by lagrangian of the form
\be
\Lg_{\rm{tot}}= \sqrt{-g}\,\left( \frac{1}{2}M_{\rm{eff}}^2 R + \L_{DE} +\L_m\right) \label{Ltot}\,,
\ee
where the subscripts $m$ and $\rm{DE}$ correspond to  matter and dark energy, and $R$ is the Ricci scalar.
The fields equations obtained by varying with respect to the metric are
\be
G_{\mu\nu}=\frac{1}{M_{\rm{eff}}^2} \left[T^m_{\mu\nu}+T^{DE}_{\mu\nu}\right]\,, \label{FE}
\ee 
where $T^m_{\mu\nu}$ and $T^{DE}_{\mu\nu}$ denote the matter and effective dark energy momentum tensors. The field equations can be always re-cast in this form by an appropriate manipulation. At this stage we have not specified the form of the dark energy Lagrangian, so eq.(\ref{Ltot}) represent a very large class of theories.
In deriving the above equation we have assumed minimal coupling between the metric and the matter fields, i.e. we are using the Jordan frame. 

The perturbation of eq.(\ref{FE}) gives
\be
\delta G_{\mu\nu}=\frac{1}{M_{\rm{eff}}^2} \left[\delta T^m_{\mu\nu}+\delta T^{DE}_{\mu\nu}\right]\,.
\label{FEP}
\ee 
Denoting with a hat the solutions of eq.(\ref{FEP}) we can manipulate it order to absorb the effects od dark energy perturbations in an appropriately defined effective gravitation coupling
\be
\delta \hat{G}_{\mu\nu}=\frac{1}{M_{\rm{eff}}^2} \left(1+\frac{\delta \hat{T}^{DE}_{\mu\nu}}{\delta \hat{T}^m_{\mu\nu}}\right)\delta \hat{T}^{m}_{\mu\nu}
\label{GeffGen}=G^{\mu\nu}_{\rm{eff}} \delta \hat{T}^{m}_{\mu\nu}%G^{(0)}+\delta\G
\ee

\section{Cosmological effects on scalar perturbations}
When studying the effects of dark energy on the background, the  energy momentum tensors are normally treated separately, but for cosmological perturbations the effects of dark energy are often modeled in terms of a modification of the general relativity (GR) equations, in terms of an effective gravitational coupling \cite{Ishak:2024jhs}.

For example, in the Newton gauge the  linearized scalar perturbations equation with respect to a Friedmann-Robertson-Walker (FRW) background takes the  form

\be
   \nabla^{2} \Psi =  \frac{a^{2} }{2 M^2_{\rm eff}} \left(\delta\rho_m + \delta \rho_{DE}\right)\,, \\  \label{P1}
\ee
where $\delta\rho_m$  are the comoving matter perturbations and  $\delta\rho_{DE}$ denotes the dark energy perturbations.
In the $\Lambda$CDM limit the above equation reduces to the  Poisson's equation in comoving coordinates, in which case $G_N=1/M^2_{\rm eff}$ and $\delta \rho_{DE}=0$.

For a given set of solutions of the cosmological perturbations equations we  can re-write eq.(\ref{P1}) as
\be
   \nabla^{2} \hat{\Psi} =   \frac{a^{2}}{2 M^2_{\rm eff}} \left(1 + \frac{\delta \hat{\rho}_{DE}}{\delta\hat{\rho}_m}\right) \delta\hat{\rho}_m = \frac{1}{2} a^{2} G^{\Psi}_{\rm{eff}} \delta\hat{\rho}_m\\ \,, \label{P1eff}
\ee
where we are denoting with a hat the solutions of the perturbations equations, and we have defined the effective gravitational coupling as 
\be
G^{\Psi}_{\rm{eff}}(\eta,x^i)=\frac{1}{M^2_{\rm eff}} \left(1 + \frac{\delta \hat{\rho}_{DE}}{\delta\hat{\rho}_m}\right)\,.\label{Peff}
\ee
The effective eq.(\ref{P1eff}) admits by construction $\hat{\psi}$ as a solution, so it provides an effective equivalent description of the system, which can always be obtained form the solutions of the perturbations equations. 
See for example \cite{DeFelice:2011hq} for an explicit calculation of $G^{\Psi}_{\rm{eff}}$, based on analytical solutions of the perturbations equations for Honrdeski theories. Note that this effective description is only valid for $\delta\rho_m \neq 0$.
Since the cosmological perturbations solutions depends on both space and time, also $G^{\Psi}_{\rm{eff}}$ is both space and time dependent.
Given the generality of the assumptions made to derive eq.(\ref{P1eff}), we can conclude that an effective Poisson equation of the form given in eq.(\ref{P1eff}), can provide a general effective description of  the effects of dark energy on structure formation.

A similar procedure can be followed in Fourier space, giving an effective Poisson's equation of the form
\be
   \frac{k^2}{a^2} \Psi_k =  -\frac{1}{2} a^{2} G^{\Psi}_{\rm{eff}}(\eta,k) \delta\rho_{m,k}\\  \label{Peffk}
\ee
where we have defined
\be
G^{\Psi}_{\rm{eff}}(\eta,k)=\frac{1}{M^2_{\rm eff}} \left(1 + \frac{\delta \rho_{DE,k}}{\delta\rho_{m,k}}\right)\,,\label{Peffk}
\ee
and we have now dropped the hat for ease of notation.
This effective description allows to make  the general prediction that the effective gravitational coupling can be both scale and time dependent, which can be useful for model independent analysis of observational data, treating $G^{\Psi}_{\rm{eff}}(t,k)$ as a phenomenological function constrained by data.

This effective approach requires to know the cosmological perturbations equations solutions to compute the effective gravitational coupling, i.e. it gives a general procedure to obtain the  effective coupling, but it does not predict its explicit analytical form in terms of langragian coefficient.
In this regard it is useful to use the effective field theory of dark energy to derive general analytical expression for $G^{\Psi}_{\rm{eff}}$, with the advantage of allowing to derive precise predictions for the interconnection with other observables \cite{Romano:2025apm,Romano:2025pcs}.

The Bardeen potentials are related by
\be
\Phi-\Psi=\frac{a^2}{M^2_{\rm{eff}}} \Pi_{DE}\,,
\ee
where $\Pi_{DE}$ is the effective scalar anisotropy potential of dark energy, from which we get
\be
\nabla^{2} (\Psi+\Phi)= \nabla^{2} \left(2\Psi+\frac{a^2}{M^2_{\rm{eff}}}\Pi_{DE}\right)=  a^{2} G^{\Psi}_{\rm{eff}} \delta\rho_m +\frac{a^2}{M^2_{\rm{eff}}} \nabla^{2} \Pi_{DE}= a^{2} G^{\Psi+\Phi}_{\rm eff} 
\delta\rho_m\,,
\ee
where we have defined
\be
G^{\Psi+\Phi}_{\rm eff}=G^{\Psi}_{\rm{eff}}+\frac{1}{M^2_{\rm{eff}}}\frac{\nabla^{2} \Pi_{DE}}{\delta\hat{\rho}_m}=\frac{1}{M^2_{\rm{eff}}}\left(1 + \frac{\delta \hat{\rho}_{DE}}{\delta\hat{\rho}_m}+\frac{\nabla^{2} \Pi_{DE}}{\delta\hat{\rho}_m}\right)\,.
\ee
%A similar approach can also be applied to the other Bardeen potential $\Phi$, and for the equation relevant for weak lensing it gives
%\be  
%\nabla^{2} (\Psi+\Phi) =  a^{2} G^{\Psi+\Phi}_{\rm eff} 
%\delta\rho_m\, \,, 
%\ee
In momentum space we obtain 
\be  
\frac{k^2}{a^2} (\Psi_k+\Phi_k) =  -a^{2} G^{\Psi+\Phi}_{\rm eff}(\eta,k) 
\delta\rho_{m,k}\, \,,
\ee
where
\be
G^{\Psi+\Phi}_{\rm eff}(\eta,k)=G^{\Psi}_{\rm{eff}}(\eta,k)-\frac{1}{M^2_{\rm{eff}}}\frac{k^2 \Pi_{DE,k}}{\delta\hat{\rho}_{m,k}}\,. %=\frac{1}{M^2_{\rm{eff}}}\left(1 + \frac{\delta \rho_{DE}}{\delta\rho_m}+\frac{\nabla^{2} \Pi_{DE}}{\delta\rho_m}\right)\,.
\ee
Examples of explicit analytical calculations of $G_{\rm{eff}}^{\Psi+\Phi}$ and $G_{\rm{eff}}^{\Psi}$ for Horndeski theories in the deep sub-horizon limit, can be found in \cite{Linder:2015rcz}.
Note that this definition of effective gravitational coupling is consistent with the scale and time dependent ansatz used in the phenomenological analysis of large scale structure data \cite{Ishak:2024jhs}.

\section{Cosmological effects on gravitational waves}
Dark energy also effects the propagation of gravitational waves, since the dark energy perturbations act as an additional effective energy-stress tensor.
The perturbations of eq.(\ref{FE}) give rise to different modifications of the GR propagation equation.
For general GWs $h_{ij}$ propagating in the $z$-direction \cite{Ezquiaga:2018btd}, the spatial part of the metric corresponding to different polarizations can be written as
 \be
 \label{Aij}
 h_{ij}=\begin{pmatrix} h_S+h_{+} & h_\times & h_{V1} \\ h_\times & h_S-h_+ & h_{V2} \\ h_{V1} & h_{V2} & h_L \end{pmatrix}\,,
 \ee
 where $h_+$ and $h_\times$ are the  tensor modes, $h_{V1,2}$ the  vector polarizations, and $h_S$   and $h_L$ the transverse and longitudinal scalar modes. % Note that the gauge has not been fully fixed in the above equation, and this can lead to GWs modes which depend on the observer frame \cite{Bonvin:2022mkw}. We will resolve this ambiguity in the next section, in which we will obtain the effective action for  gauge invariant GWs.  \cite{Nishizawa:2017nef}
 Ignoring the interaction of GWs with matter, the free action in a FRW background is given by \cite{Romano:2024apw,Gubitosi:2012hu}
 \be
\frac{a^2 M^2_{\rm{eff}}}{c_{ij}^2}\Big[ h'^2_A-c_{ij}^2 (\nabla h_{ij})^2\Big]
\ee
giving the propagation equation 
\be
h_{ij}''+2  \h\Big(1-\frac{c_{ij}'}{\h c_{ij}}+\frac{M_{\rm{eff}}'}{\h M_{\rm{eff}}}\Big) h'_{ij}-c_{ij}^2 \nabla^2 h_{ij}=0  \,,\label{hctO}
\ee
where $ \h=a'/a$.
Including the interaction terms we get
\be
h_{ij}''+2  \h\Big(1-\frac{c_{ij}'}{\h c_{ij}}+\frac{M_{\rm{eff}}'}{\h M_{\rm{eff}}}\Big) h'_{ij}-c_{ij}^2 \nabla^2 h_{ij}=\frac{a^2}{M^2_{\rm{eff}}}(\delta T^m_{ij}+\delta T^{DE}_{ij})=a^2 G^{ij}_{\rm{eff}} \delta T^m_{ij}  \,,\label{h1}
\ee
where we have defined the effective gravitational coupling for the different GWs polarizations as
\be
G^{ij}_{\rm{eff}}=\frac{1}{M^2_{\rm eff}} \left(1 + \frac{\delta T^{DE}_{ij}}{\delta T^m_{ij}}\right)\,.\label{Pijeff}
\ee
Note that the indices $ij$ in $G^{ij}_{\rm{eff}}$ are not tensorial, but have only a labeling function, and are not summed in the source term or in any other equation.
The full first order effective action corresponding to eq.(\ref{h1}), including the effects of interaction with matter, is 
\be
\Lg^{(1)}_h=\frac{a^2 M^2_{\rm{eff}}}{c_{ij}^2}\Big[ h'^2_{ij}-c_{ij}^2 (\nabla h_{ij})^2\Big]+ a^4 G^{ij}_{\rm{eff}} h^{ij}\delta T^m_{ij}\,. 
\ee

Note that since cosmological perturbations are space and time dependent, in general also $G^{ij}_{\rm{eff}}(\eta,x)$ is expected to be  also space and time dependent.
Contrary to GR, in a general modified gravity theory the vector and scalar degrees of freedom cannot be gauged away, because even in absence of matter perturbations, the DE effective energy tensor will act as source for these  \cite{Flanagan:2005yc}.
Note that the GWs propagation equation can be always re-cast in the form in eq.(\ref{h1}), by absorbing in $\delta T^{DE}_{ij}$ any  modified gravity effect, including for example graviton mass terms \cite{deRham:2014zqa}.
An alternative effective approach consists in encoding the effects of the source term in a frequency-time-polarization dependent effective speed \cite{Romano:2022jeh}, which can be convenient for the calculation of the effects on gravity modification on the GW-EMW distance ratio.

In momentum space we can obtain a similar result
\be
h_{ij,k}''+2  \h\Big(1-\frac{c_{ij}'}{\h c_{ij}}+\frac{M_{\rm{eff}}'}{\h M_{\rm{eff}}}\Big) h'_{ij,k}+ c_{ij}^2 k^2 h_{ij,k}=\frac{a^2}{M^2_{\rm{eff}}}(\delta T^m_{ij,k}+\delta T^{DE}_{ij,k})=a^2 G^{ij}_{\rm{eff}} \delta T^m_{ij,k}  \,,\label{h1k}
\ee
where we have defined the scale dependent effective gravitational coupling for the different GWs polarizations as

\be
G^{ij}_{\rm{eff}}(\eta,k)=\frac{1}{M^2_{\rm eff}} \left(1 + \frac{\delta T^{DE}_{ij,k}}{\delta T^m_{ij,k}}\right)\,.\label{Pijeff}
\ee

\section{General relativity limit}
When the dark energy perturbations are absent, i.e. when ${\delta T^{DE}_{ij}}=\delta\rho_{DE}=0$, the effective gravitational coupling for scalar perturbations and GWs are the same
\be
G^{\Psi+\phi}_{\rm{eff}}=G^{\Psi}_{\rm{eff}}=G^{ij}_{\rm{eff}}=\frac{1}{M^2_{\rm eff}} \,,
\ee
and if the effective Planck mass is not running we obtain the GR limit
\be
G^{\Psi+\phi}_{\rm{eff}}=G^{\Psi}_{\rm{eff}}=G^{ij}_{\rm{eff}}=\frac{1}{M^2_{\rm p}} \,,
\ee
where $M_p$ is the Planck mass.
Note that in general each GWs polarization and the different scalar perturbations have different effective gravitational couplings, and none of them is simply given by $1/{M^2_{\rm eff}}$. 

\section{Local negative effective gravitational coupling}
From eq.(\ref{Peff}), we can see that the effective gravitational coupling can be negative when
\be
\frac{\delta \rho_{DE}}{\delta\rho_m}<-1\,.
\ee
 Assuming a positive matter density perturbation $\delta\rho_m$, the effective gravitational coupling can be locally negative if the dark energy density perturbations $\delta\rho_{DE}$ are negative. This can happen locally in regions of space where $\delta\rho_{DE}(\eta,x^i)<-\delta\rho_{m}(\eta,x^i)$, or in terms of the density contrast, when $\delta_{DE}(\eta,x^i)<-\delta_{m}(\eta,x^i)$, i.e. a sufficiently large local dark energy under-density produces a negative effective gravitational coupling. This is a local effect which depends on the local fluctuations of dark energy perturbations, and does not require negative energies, since the total dark energy density can be positive even if $\delta_{DE}<0$. Since the minimum value of the density contrast is by definition $-1$, the effective gravitational coupling can be negative only for $\delta_m<1$. 

\section{Phantom dark energy induced structure suppression}

\begin{figure}[h]
\includegraphics[width=\columnwidth]{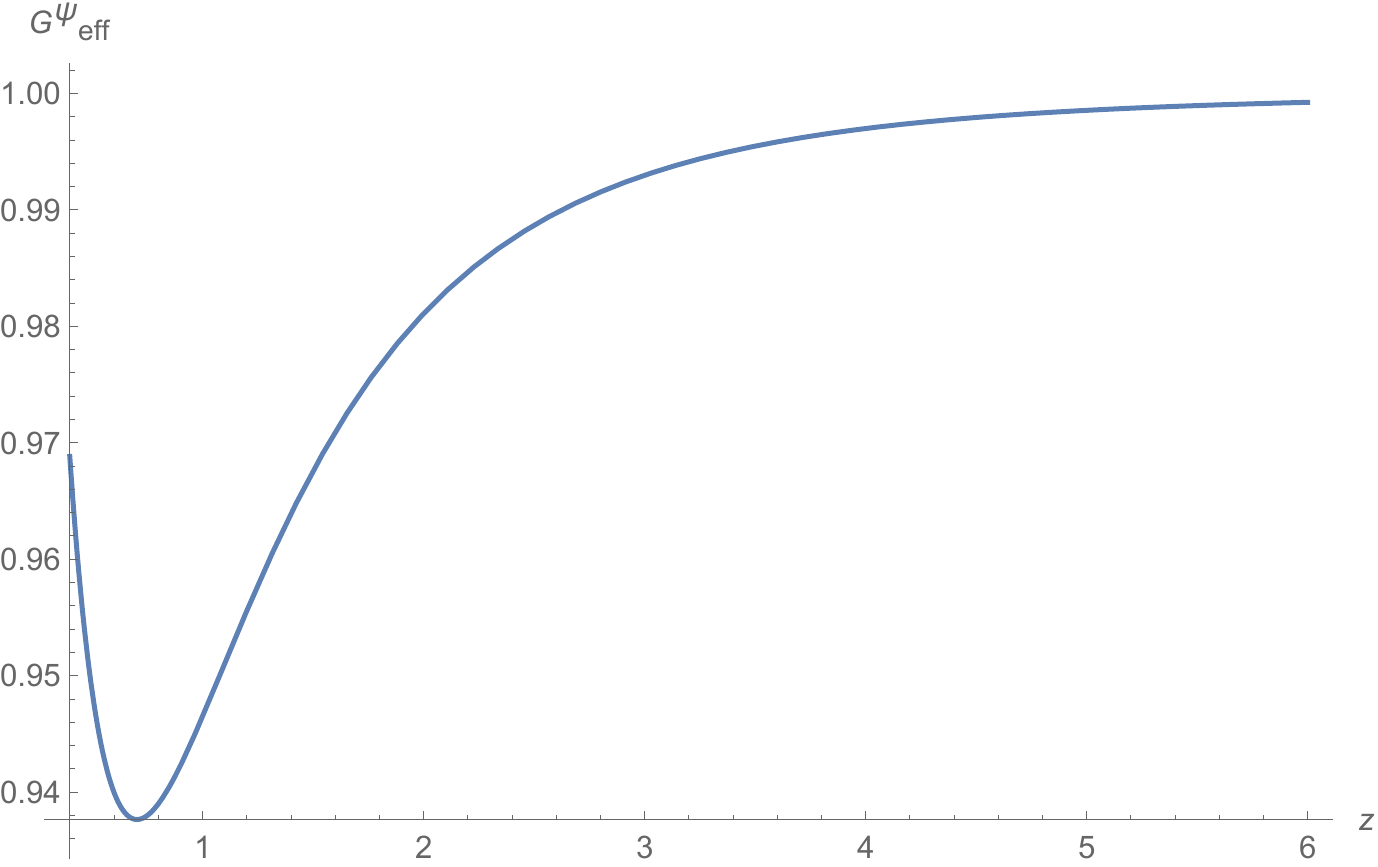}
\caption{The effective gravitational coupling corresponding to the dark energy equation of state parameters estimated by DESI is plotted as function of redshift, in units of $1/M^2_{\rm eff}$. The plot corresponds to $w(a)=w_0+w_a(1-a)$, with $w_0=-0.827$ and $w_a=-0.75$, which are the best fit values of the DESI+CMB
 +PantheonPlus analysis \cite{DESI:2024mwx}.}
\label{fig:GeffDESI}
 \end{figure}

Beside the local effect mentioned above, there can also be a cosmological effect when matter and dark energy perturbations are adiabatic, i.e. when $\delta\rho_{DE}/\rho'_{DE}=\delta\rho_{m}/\rho'_{m}$, in which case eq.(\ref{Peff}) gives

\be
G^{\Psi}_{\rm{eff}}=\frac{1}{M^2_{\rm eff}} \left(1 + \frac{\rho'_{DE}}{\rho'_{m}} \right)=\frac{1}{M^2_{\rm eff}} \left[1+(1+w_{DE})\frac{\rho_{DE}}{\rho_m}\right]\,,\label{Pad}
\ee
where we have used the continuity equation for dark energy, $\rho_{DE}'=-3/a\Big[1+w_{DE}(a)\Big]\rho_{DE}$, and for matter, $\rho_{m}'=-3/a \, \rho_{m}$. 
At late time, when dark energy is dominating, and if $w_{DE}<-1$, the effective gravitational coupling can be reduced with respect to its early time value $1/M^2_{\rm eff}$, corresponding to when dark energy is subdominat, i.e $\rho_{DE}/\rho_{m}\approx 0$.
Quite remarkably, the recent analysis of the Dark Energy Spectroscopic Instrument (DESI)  \cite{DESI:2024mwx} shows a preference for dynamical dark energy with $w<-1$, and at low redshift  there is also an evidence of  suppression of structure \cite{Chen:2024vuf}. 
In fig.(\ref{fig:GeffDESI}) it is shown the redshift dependence of the effective gravitational coupling corresponding to the dark energy equation of state parameters estimated by DESI \cite{DESI:2024mwx}, confirming that at low redshift there can be a reduction  compared to the early time value, and providing a possible explanation for the observed low redshift structure suppression \cite{Chen:2024vuf}. This shows that the observed structure suppression at low redshift could be another manifestation of the background dark energy equation of state, providing further independent evidence for this type of dynamical dark energy.

\section{Conclusions}
We have shown that the effects of dark energy can be encoded in a momentum, space and polarization dependent effective  gravitational coupling, and that this is not just related to the effective Planck mass appearing in the action as a coefficient of the Ricci's scalar.
The generality of this effective description of the effects of dark energy on cosmological perturbations gives a solid theoretical justification for the phenomenological approaches often used in the analysis of cosmological observations, and allows to understand the mapping between phenomenological analysis results \cite{Ishak:2024jhs,Nishizawa:2017nef} and theoretical models.
This effective description naturally predicts that the gravitational wave radiation emitted by binary systems can have an additional frequency and polarization dependence, and it will be interesting to test this prediction with future GWs observations.

For scalar perturbations the scale dependency of the effective gravitational coupling allows to provide an effective description for the screening mechanism, i.e. the suppression of the effects of dark energy on small scales, which is often necessary  in order to pass local gravitational tests. 
The scalar perturbations effective gravitational coupling can be locally negative, in regions with local dark energy under-densities.
For adiabatic perturbations the effective gravitational coupling for scalar perturbations can be negative for phantom dark energy, providing a possible explanation for the observed structure suppression at low redshift \cite{Chen:2024vuf}. 

While in this paper we have not considered specific dark energy models, the generality of the assumptions have allowed to derive a general model independent theoretical framework for the study of the effects of dark energy perturbations on cosmological perturbations which can be applied to a large class of theories, and can be convenient for model independent analysis of observational data.

\section{Acknowledgements}
This work was supported by the UDEA project No. 2023-63330.

\appendix

\section{Physical implications for scalar and tensor perturbations}
We have shown that the effects of dark energy on cosmological perturbations can be modeled by a momentum, time and polarization dependent effective gravitational coupling.
A direct implication of this general result is that the gravitational energy  radiated by a binary system can acquire an additional dependence on the GWs frequency and polarization. This is due to the fact that the calculation of the radiated power involves the solution of the wave equation for a binary coalescence, and the momentum and polarization dependence of the effective gravitational coupling will induce new effects.
Formally we can write the solution of the GWs equation as
\be
h_{ij}=\Box^{-1}\left({G^{ij}_{\rm{eff}}} \delta T^{m}_{ij}\right) \,,
\ee
which differs from the GR case in which $G^{ij}_{\rm{eff}}$ is a constant. Since $G^{ij}_{\rm{eff}}$ is polarization and space  dependent, different polarizations can have a different radiated power. Writing the wave equation in momentum space, the momentum dependence of $G^{ij}_{\rm{eff}}$ would also induce a new additional frequency dependency. %Here we just mention why there should be a physical effect on the radiated power due to the effective gravitational coupling momentum-polarization dependence, and we leave the derivation of a generalized quadrupole formula for the radiated power to a future work.
In a general modified gravity theory there can be GWs of scalar and vector polarization \cite{Ezquiaga:2018btd}, which cannot be gauged away, contrary to GR. %Performing a scalar-vector-tensor decomposition of the metric perturbations, for the tensor part, i.e. the  transverse and traceless (TT) piece \cite{Flanagan:2005yc},
At leading order we have 
\begin{align}\label{Solh}
    &h_{ij}=\frac{2G^{ij}_{\rm{eff}}}{r}\Ddot{Q}_{ij},
\end{align}
where $Q$ is the traceless mass quadrupole moment.
Accounting for the time and polarization dependence, but neglecting momentum dependence of the effective gravitational coupling, following a similar procedure as the one outlined in \cite{Wolf:2019hun} we get the following expression for the radiated power of different  polarizations
\be
\label{Pgw1}
    P_{ij} = \frac{G^{ij}_{\rm{eff}}}{5}\left[  \langle \dddot{Q}_{ij} \dddot{Q}^{ij} \rangle + \frac{2\dot{G}^{ij}_{\rm{eff}}}{G^{ij}_{\rm{eff}}} \langle \Ddot{Q}_{ij} \dddot{Q}^{ij}\rangle + \left(\frac{\dot{G}^{ij}_{\rm{eff}}}{G^{ij}_{\rm{eff}}}\right)^2 \langle \Ddot{Q}_{ij} \Ddot{Q}^{ij} \rangle  \right].
\ee
Since $G^{ij}_{\rm{eff}}$ depends on the GWs polarizations, the power radiated can also be polarization dependent, due to the potentially  different way in which different polarizations interact with dark energy perturbations. We leave to a future work the derivation of a generalized quadrupole formula, including the frequency dependency due to the scale dependency of the effective gravitational coupling.

\section{Encoding higher order perturbations effects in the effective coupling}

We have so far considered linear order metric perturbations, and at higher order $p$ other  gauge invariant variables ${h}^{(p)}_{\mu\nu},\Psi^{(p)},\Phi^{(p)}$ can be defined \cite{Mollerach:2003nq,DeLuca:2019ufz,Chang:2020iji}. 
Denoting as $\hat{\Psi}^{(p)}$ and $\delta\hat{\rho}^{(p)}_m$ the solutions of the perturbations equations of order p
\be
E^{(p)}[\delta g^{(p)}_{\mu\nu},\delta\T^{(p)}_{\mu\nu}]=0\,,
\ee
we can define the effective dark energy perturbations $\delta\rho^{(p),\rm{eff}}_{DE}$ by
\be
   \nabla^{2} \hat{\Psi}^{(p)}  =  a^{2} \frac{1}{M^2_{\rm{eff}}} \left[  \delta{\hat{\rho}}^{(p)}_m+\delta\rho^{(p),\rm{eff}}_{DE}\right]\,,\label{rhoDEeff}
\ee
which shows that the solution $\Psi^{(p)}$ of order p can be obtained as solution of the effective eq.(\ref{rhoDEeff}) for an appropriate choice of $\delta\rho^{(p),\rm{eff}}_{DE}$. %,
%from which we obtain eq.(\ref{Psip}).
%This shows that any solution $\hat{\Psi}^{(p)}$ can be obtained as a solution of the effective eq.(\ref{Psip}), by appropriately defining $\delta\rho^{(p),\rm{eff}}_{DE}$ with eq.(\ref{rhoDEeff}).
Following a method similar to the one used to derive eq.(\ref{P1eff}) we can then obtain the effective Poisson's equation of order p  
\be
   \nabla^{2} \Psi^{(p)}  =  a^{2} G^{\Psi(p)}_{\rm{eff}} \delta{\rho}^{(p)}_m\,,\\  \label{Psip}
\ee
where we have defined the effective gravitational coupling of order p as 
\be
G^{\Psi(p)}_{\rm{eff}}=\frac{1}{M^2_{\rm eff}} \left(1 + \frac{\delta\rho^{(p)}_{DE}}{\delta\rho^{(p)}_m}\right)\,.\label{Peffp}
\ee
Summing the equations at different orders in perturbations we get
\be
   \nabla^{2} \Psi^{(N)}  =  a^{2} G^{\Psi(N)}_{\rm{eff}} \delta\rho^{(N)}_m\,,\\ \label{PN}
\ee
where we have defined the summed Bardeen potential $\Psi^{(N)}$ and energy-tensor perturbations
\be
\Psi^{(N)}=\sum^N_{p=1} \Psi^{(p)} \quad,\quad \delta\rho^{(N)}_m =\sum^N_{p=1}  \delta\rho^{(p)}_m\label{PsiN} \,,
\ee
and the effective gravitational coupling is
\be
G^{\Psi(N)}_{\rm{eff}}=\frac{1}{\delta\rho^{(N)}_m}\sum^N_{p=1} G^{\Psi(p)}_{\rm{eff}} \delta\rho^{(p)}_m \,.
\ee
%The effective equation \ref{PsiN} can be obtained from  the effective action
This shows that the sum $\Psi^{(N)}$ of the solutions of cosmological perturbations equations at different orders can be obtained as a solution of eq.(\ref{PN}), making it a convenient tool for the phenomenological study of the effects of dark energy.
%Note that the scalar perturbations equation of order p can be always re-cast in the form given in eq.(\ref{Psip}). 

For GWs we can follow a similar effective approach. First it is convenient to note that eq.(\ref{hctO}) can be  written in terms of the D'Alambert operator in curved space as
\be
\Box h^{(1)}=0 \,,
\ee
defined by the effective  metric \cite{Romano:2023bzn,Romano:2022jeh},
\be
ds_{eff}^2= M^2_{\rm{eff}} a^2 \Big[c_{ij} d\eta^2-\frac{\delta_{ij}}{c_{ij}}dx^idx^j\Big] \label{geffzeta} \,,
\ee
where we are denoting explicitly the order of perturbations.
Using a similar procedure to the one outlined above for $\delta\rho^{(p),\rm{eff}}_{DE}$, we can define $\delta T^{DE(p)}_{ij,\rm{eff}}$ by 
\be
\Box \hat{h}^{(p)}_{ij}=\frac{a^2}{M^2_{\rm{eff}}}\left[\delta \hat{T}^{m(p)}_{ij}+\delta T^{DE(p)}_{ij,\rm{eff}}\right] \,,\label{deltaTeffp}
\ee
where $\hat{h}$ and $\delta \hat{T}^{m(p)}_{ij}$ are the solutions of the cosmological perturbations equations of order p.
This shows that any solution $\hat{h}^{(p)}$ can be obtained as a solution of the effective eq.(\ref{deltaTeffp}), by appropriately defining $\delta T^{DE(p)}_{ij,\rm{eff}}$.
From Eq.(\ref{deltaTeffp}) we can then obtain the effective equation in terms of the effective gravitational coupling
\be
\Box h^{(p)}_{ij}=a^2{G^{ij(p)}_{\rm{eff}}} \delta T^{m(p)}_{ij} \label{hp} \,,
\ee
where we have defined the effective gravitational coupling for the different GWs polarizations as
\be
G^{ij(p)}_{\rm{eff}}=\frac{1}{M^2_{\rm eff}} \left(1 + \frac{\delta T^{DE(p)}_{ij,\rm{eff}}}{\delta T^m_{ij(p)}}\right)\,.\label{Pijmeff}
\ee
%This can always be achieved by appropriately defining the effective perturbed energy-stress tensor $\delta T_{ij}^{DE(p)}$, even if the d'Alambert operator does not appear explicitly in the perturbed equations, by adding it on both sides of the equations. % This effective equation allows to interpret GWs at any order in perturbations as the solutions of a wave equation with an appropriately defined source term.
The d'Alambert operator is defined w.r.t. to  the same effective metric in all these equations, so they can be summed to give
\be
\Box h^{(N)}_{ij}=a^2 {G^{ij(N)}_{\rm{eff}}}\delta T^{m(N)}_{ij}\label{hTN}  \,,
\ee
where we have used the  is linearity of the d'Alambert operator, and  defined the summed GWs and energy-tensor perturbations
\be
h^{(N)}_{ij}=\sum^N_{p=1} h^{(p)}_{ij} \quad,\quad T^{(N)m}_{ij} =\sum^N_{p=1} \delta T^{(p)m}_{ij} \,,
\ee
and 
\be
G^{ij(N)}_{\rm{eff}}=\frac{1}{\delta T^{(N)m}_{ij}}\sum^N_{p=1}   {G^{ij(p)}_{\rm{eff}}} \delta T^{(p)m}_{ij}  \label{GijN}\,.
\ee
The quantities $h^{(N)}_{\mu\nu}$ are the physically observable GWs, obtained by summing  the contributions from different perturbations orders. We have shown that the sum $h^{(N)}_{\mu\nu}$ of the solutions of the perturbations equations at different orders can be obtained as the solution of eq.(\ref{hTN}) with an effective coupling given by eq.(\ref{GijN}). This implies that eq.(\ref{hTN}) can be used to model phenomenologically the effects of dark energy, including the non linear effects.
The effective action corresponding to eq.(\ref{hTN}) is
\be
\Lg^{(N)}_h=\frac{a^2 M^2_{\rm{eff}}}{c_{ij}^2}\Big[ {h_{ij}^{(N)'}}^2-c_{ij}^2 {\left(\nabla h_{ij}^{(N)}\right)}^2\Big]+a^4 G^{ij(N)}_{\rm{eff}} h^{ij{(N)}}\delta T^{(N)m}_{ij}\,.
\ee

\bibliographystyle{apsrev4-1}
\bibliography{Bib}
\end{document}